# Is the interstellar object 3I/ATLAS alien technology?

**Adam Hibberd,[1] Adam Crowl,[1] Abraham Loeb[2]**

[1]Initiative for Interstellar Studies (i4is), 27/29 South Lambeth Road London, SW8 1SZ, United Kingdom

[2]Astronomy Department, Harvard University, 60 Garden Street, Cambridge MA 02138, USA

## Abstract

At this early stage of its passage through our Solar System, 3I/ATLAS, the recently discovered interstellar interloper, has displayed various anomalous characteristics, determined from photometric and astrometric observations. As largely a pedagogical exercise, in this paper we present additional analysis into the astrodynamics of 3I/ATLAS, and hypothesize that this object could be technological, and possibly hostile as would be expected from the 'Dark Forest' resolution to the 'Fermi Paradox'. We show that 3I/ATLAS approaches surprisingly close to Venus, Mars and Jupiter, with a probability of $\lesssim 0.005\%$. Furthermore the low retrograde tilt of 3I/ATLAS's orbital plane to the ecliptic offers various benefits to an Extra-terrestrial Intelligence (ETI), since it allows the object access to our planet with relative impunity. The eclipse by the Sun from Earth of 3I/ATLAS at perihelion, would allow it to conduct a clandestine reverse Solar Oberth Manoeuvre, an optimal high-thrust strategy for interstellar spacecraft to brake and stay bound to the Sun. An optimal intercept of Earth would entail an arrival in late November/early December of 2025, and also, a non-gravitational acceleration of $\sim 5.9 \times 10^{-5}$ au day$^{-2}$, normalized at 1 au from the Sun, would indicate an intent to intercept the planet Jupiter, not far off its path, and a strategy to rendezvous with it after perihelion.

**Correspondence:** Adam Hibberd, Initiative for Interstellar Studies (i4is), 27/29 South Lambeth Road London, SW8 1SZ, United Kingdom, Email adam.hibberd@i4is.org

## Introduction

This paper is contingent on a remarkable but, as we shall show, testable hypothesis, to which the authors do not necessarily ascribe, yet is certainly worthy of an analysis and a report, for two reasons:

1. The consequences, should the hypothesis turn out to be correct, could potentially be dire for humanity, and would possibly require defensive measures to be undertaken (though these might prove futile).
2. The hypothesis is an interesting exercise in its own right, and is fun to pursue, irrespective of its likely validity.

The hypothesis in question is that the recent interstellar visitor to our Solar System, 3I/ATLAS[1-10] is a technological artifact, and furthermore has active intelligence. If this is the case, then two possibilities follow: first that its intentions are entirely benign and second they are malign, or somewhere in-between.

To address the extreme cases in turn, in the first case, humanity need do nothing save await the arrival of this intelligence with open arms. It is the second eventuality which is of most concern, and according to the so-called 'Dark Forest' resolution to the 'Fermi Paradox', would be more likely, as it would neatly explain the singular lack of success of the SETI initiative to-date.[11]

Discovered on 1$^{st}$ July 2025, by the *Asteroid Terrestrial-impact Last Alert System*, 3I/ATLAS is, as its designation indicates, the latest interstellar object (or interloper) to be discovered passing through our Solar System. The first, 1I/'Oumuamua, detected in 2017, was only visible for a period of 2 months, though various anomalous features of this object have yet to be clarified.[12,13] Despite this, there are still very entrenched opinions on the subject in the scientific community as Eladi, Tenenbaum and Loeb submitted to "Psychological Review" on July 9$^{th}$, 2025.[14]

Perhaps one of the most puzzling observations is the presence of a statistically significant 'non-gravitational' acceleration (i.e. $4.92 \pm 0.16) \times 10^{-6}$ m s$^{-1}$ (normalized to a distance of 1 au from the Sun),[15] despite there being no evidence of cometary outgassing from 1I/'Oumuamua,[16] the most likely cause of non-gravitational accelerations of this kind.

The discourse on whether the object 1I/Oumuamua was artifical, i.e. the non-gravitational force was actually solar radiation pressure (SRP) on extremely thin photonic (solar) sails,[17] has been mired in bitter controversy. Yet nevertheless it seems to the authors as a hypothesis perfectly worthy of pursuing, in a similar fashion to the hypothesis proposed in this paper, and the consequences derived can then be rejected or accepted accordingly.

We employ 'OITS' (or 'Optimum Interplanetary Trajectory Software'), the interplanetary mission design tool, to investigate the likelihood and implications of 3I/ATLAS being an alien spacecraft, with high and/or low thrust manoeuvrability. For further information regarding OITS, proceed to Hibberd[18,19] and Hibberd et al.[20] For this analysis, two possible Non-Linear Problem (NLP) solver options are available, namely NOMAD[21]or MIDACO.[22-24] The efficacy of this software has been proven for a variety of applications, for example for previous interstellar objects, such as 1I/'Oumuamua and 2I/Borisov as well as for terrestrial planets.[20,25-30]

Table 1 summarises the different factors which support our hypothesis. These are each addressed in turn below.

**Table 1** Considerations which support the hypothesis that 3I/ATLAS is technological

| Evidence | Description | Details |
|---|---|---|
| 1 | 3I/ATLAS orbital plane lies virtually in the Ecliptic, though retrogade, $i = 175.11^{\circ}$ | p ~ 0.2% |
| 2 | 3I/ATLAS is too large to be an asteroid | $p \lesssim 10^{-6} \times 1I$ |
| 3 | 3I/ATLAS shows no evidence of cometary outgassing | No spectral signs |
| 4 | 3I/ATLAS approaches unusually close to Venus, Mars and Jupiter | p ~ 0.005 % |
| 5 | 3I/ATLAS achieves perihelion on the opposite side of the Sun to Earth | p ~ 7 % |
| 6 | The optimal point to do a reverse Solar Oberth and stay bound to the Sun is at perihelion | Refer to Figure |
| 7 | 3I/ATLAS's incoming radiant made it hard to detect sooner | |
| 8 | The ΔV needed to intercept Jupiter is small | Refer to Figure |
| 9 | The ΔV needed to intercept Mars is small | Refer to Figure |

Taking the $1^{st}$ row in Table 1, it appears that besides the fact it is clearly on a hyperbolic trajectory, that possesses a non-zero speed at an infinite distance from the Sun of ~ 60 km $s^{-1}$; there is a further extremely unusual feature of 3I/ATLAS's trajectory which is that its orbital plane is tilted only slightly from the ecliptic (~ $5^{\circ}$), and is retrograde. This means attempts by humanity to intercept it, or even more difficult rendezvous with it, are extremely challenging if not impossible with chemical rockets, yet nonetheless, as we shall see, allows 3I/ATLAS to intercept certain key target planets with relative ease. Furthermore, a low ecliptic tilt at a distance from the Sun, would enable an ETI, through astrometric measurements, to determine the orbits and masses of the Solar System planets, allowing it to prepare an optimal approach strategy to the Solar System. The likelihood for such a perfect alignment of the orbital angular momentum vector around the Sun for Earth and 3I/ATLAS is $\pi(5^{\circ}/57^{\circ})^2/(4\pi) = 2\times10^{-3}$.

Missions to 3I/ATLAS would have been much easier had the interstellar object been travelling in the ecliptic plane prograde. Figure 1 illustrates opportunities to 3I/ATLAS expire by the year end, and even then would only be achievable for a SpaceX Starship refuelled in low Earth orbit (LEO), and a spacecraft payload with nuclear thermal propulsion (NTP) (thus $C_3 < 1500$ km$^2$ s$^{-2}$ for this plot[1], constitutes an extremely challenging launcher requirement).

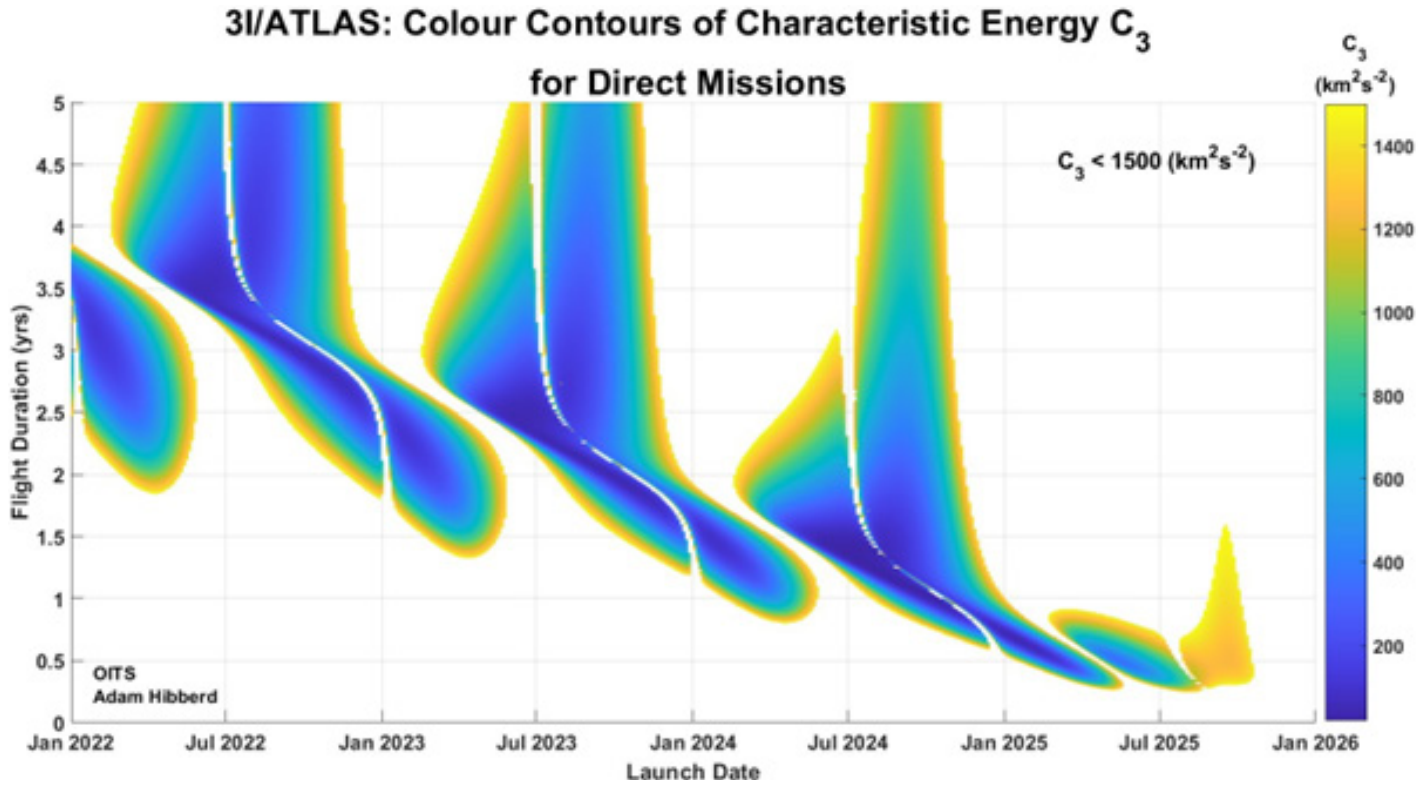

**Figure 1** Pork Chop plot detail for flyby missions to 3I/ATLAS, with $C_3 < 1500$ km$^2$ s$^{-2}$.

$2^{nd}$ in Table 1, it is clear, as-of-writing, that the true nature of 3I/ATLAS is somewhat ambiguous. Apart from the hypothesis proposed already, two distinct yet natural incarnations present themselves for this object:

1. It is an asteroid, in which case, assuming a standard albedo of 0.05, the object must be around 20 km in diameter.[6,7,9]
2. It is a comet, in which case the object would be surrounded by a fuzzy coma, with a much smaller nucleus.[6]

Both of these natural explanations present difficulties, however.

In the first case, because the prevalence of interstellar objects of size 20 km should be much lower, by many orders of magnitude, than that of objects the size of 1I/'Oumuamua (which was 2 orders of magnitude smaller than 3I/ATLAS), this then implies the visit into our Solar System of 3I/ATLAS should be an exceedingly low probability.[6]

In the second case, there has been to-date absolutely no sign from spectroscopic analysis of cometary activity on 3I/ATLAS. Such activity would imply a much smaller nucleus and allow 3I/ATLAS to be drawn from a much larger interstellar population. We await with anticipation further observations

[1] $C_3$ is known as the "Characteristic Energy" at Launch and is the square of the Earth hyperbolic excess speed on escaping the Earth's gravitational sphere of influence

of 3I/ATLAS which should clarify the situation. The fuzz observed around 3I/ATLAS is inconclusive given the motion of the object and the inevitable smearing of the image over the exposure time.[8,9]

$4^{th}$ in Table 1, we find that 3I/ATLAS approaches particularly close to Venus, Mars and Jupiter during its visit to our Solar System, refer to Table 2. In the following analysis we assume that 3I/ATLAS is on its current orbit, but vary the time-of-entry into the Solar System (or equivalently the time of perihelion), assuming 3I/ATLAS could have come at any time into the Solar System, and happened to do so such that it came within the observed closest approaches of Venus, Mars and Jupiter. The probability of this is 0.005%

We further assume that the closest approach of 3I/ATLAS to the planet is entirely a consequence of its difference in heliocentric longitude. In practice, this will not actually be the case, and this constitutes an UPPER BOUND on probability.

The $6^{th}$ column of Table 2 provides the degree of misalignment in longitude between the planet in question and the interstellar object, at its closest approach. As the orbits of the planets are nearly circular, then if this longitudinal difference is say X degrees, then the probability of the planet lying within X degrees longitude of 3I/ATLAS is:

$$P = \frac{2(X^{\circ})}{360} \qquad (1)$$

**Table 2** Pertinent Parameters concerning the alignment of 3I/ATLAS with Venus, Mars and Jupiter.

| Planet | Semi major axis (au) | Closest Approach of 3I/ATLAS (au) | Min. Poss. Closest Approach of 3I/ATLAS (au) | Max. Poss. Closest Approach of 3I/ATLAS (au) | Long. Error at Actual Closest Approach | Prob. Observed Long. Error |
|---|---|---|---|---|---|---|
| Venus | 0.723 | 0.65 | 0.627 | 2.073 | 9.95° | 5.53% |
| Mars | 1.524 | 0.19 | 0 | 2.874 | 7.14° | 3.97% |
| Jupiter | 5.203 | 0.36 | 0 | 6.553 | 3.96° | 2.20% |
| Overall Probability | | | | | | 0.005% |

The final column is a calculation of this probability in %. The overall probability of ALL 3 planets aligning in this way is the product of these 3 values and amounts to $\lesssim 0.005\%$.

The $5^{th}$ row of Table 1 shows that, at its perihelion on $29^{th}$ October 2025, when it reaches 1.35 au from the Sun, 3I/ATLAS will be totally obscured from the Earth by the Sun. If we assume this obscuration occurs within a solar elongation of 30°, then it is straight forward to calculate the likelihood of this alignment with the Sun and Earth as ~ 7%. But why should such a celestial alignment be indicative of intelligence?

Referring to the $6^{th}$ row of Table 1, we find a possible motivation, since the optimal braking strategy to stay bound to the Sun for high thrust propulsion is a 'reverse Solar Oberth', where all the thrust is imparted at perihelion. Thus, any manoeuvres of this kind would be obscured from Earth observation, allowing a surprise arrival on Earth to be conducted.

The $7^{th}$ row in Table 1 reveals a curious feature of 3I/ATLAS's apparent direction of origin, in that 3I/ATLAS's incoming radiant to the Solar System was from the direction of the Galactic Centre, a particularly bright region, which, as has been noted elsewhere, made the object particularly difficult to discern by Earth-based telescopes, in turn rendering it less conducive to early detection.

The relevance of this is that had the object indeed been discovered earlier, then there would have been some possibility that humanity could have mounted an intercept mission, a recourse that was out-of-the-question by the time 3I/ATLAS was actually detected. Figure 2 shows the optimal intercept trajectory for a mission to 3I/ATLAS, with optimal launch over a year ago. See also Figure 1.

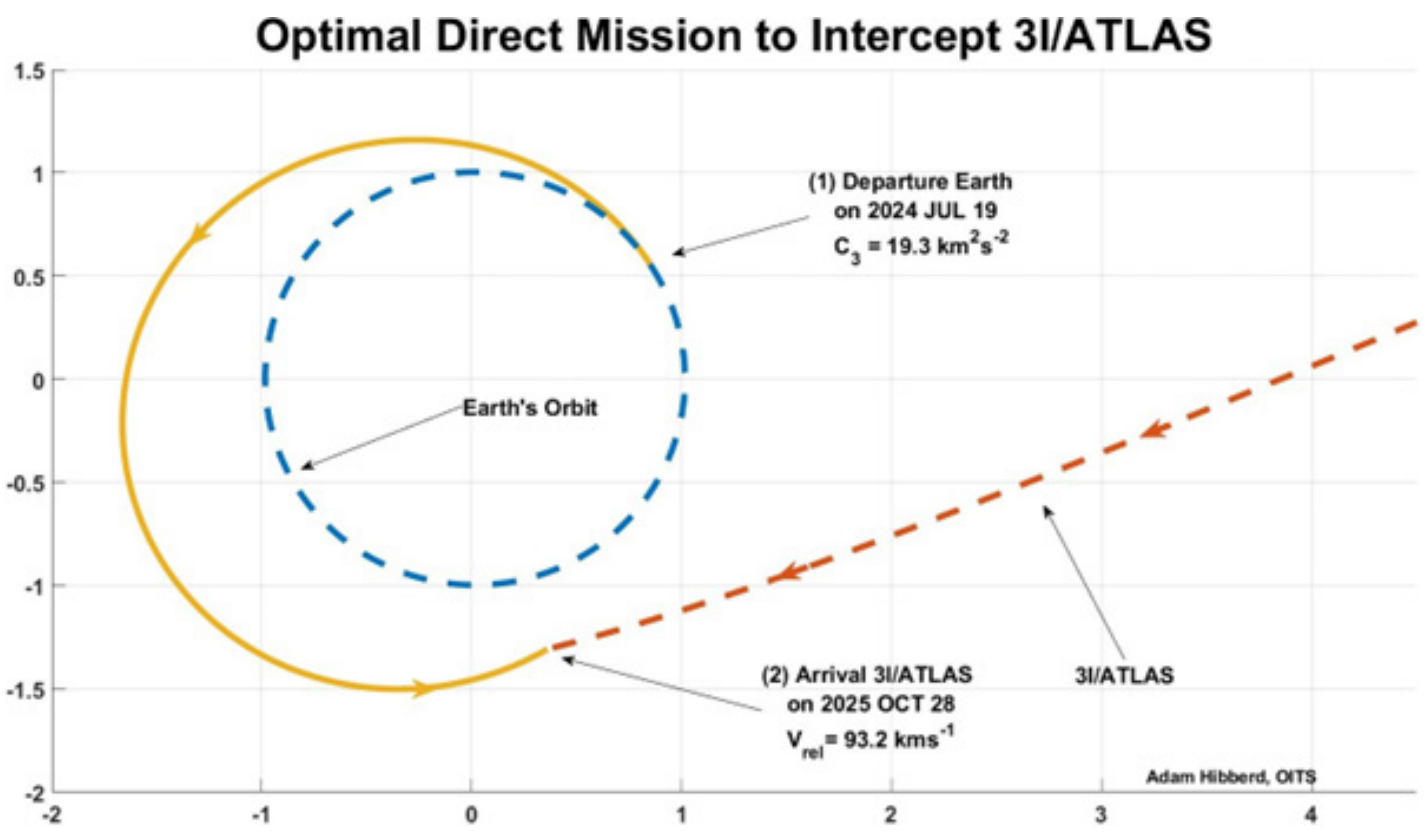

**Figure 2** Optimal trajectory to intercept 3I/ATLAS with launch date on 2024 JUL 09, a full year earlier than the discovery date of this interstellar object.

## Possible strategies and motivations

3I/ATLAS has already passed close to Pluto (~ 5.1 au) and in the future it will come very close to the inner planets Venus (~ 0.65 au) and Mars (~ 0.19 au) and Jupiter (~ 0.36 au) (see Figure 3). As a consequence, the ΔV needed by 3I/ATLAS to either (a) intercept any of these planets or (b) send probes to them, is low (see Figure 4). Mercury and Earth are exceptions, though this makes sense if 3I/ATLAS had narrowed its intentions to planets in the Sun's habitable zone, the reason for keeping its distance from Earth shall be elucidated below.

Thus, for Venus, the ΔV for intercept is < 5 km $s^{-1}$ before April 2025 and remains < 10 km $s^{-1}$ until the end of July 2025. Similarly for Mars the ΔV also stays below 5 km $s^{-1}$ until the end of July 2025. For Jupiter the intercept ΔV is low (< 5 kms$^{-1}$) all the way up to November 2025. For Earth the intercept ΔV is always above 5 km s$^{-1}$. In all cases it can be seen, as would be expected, the sooner the delivery of this ΔV, the lower its magnitude. But what does this imply?

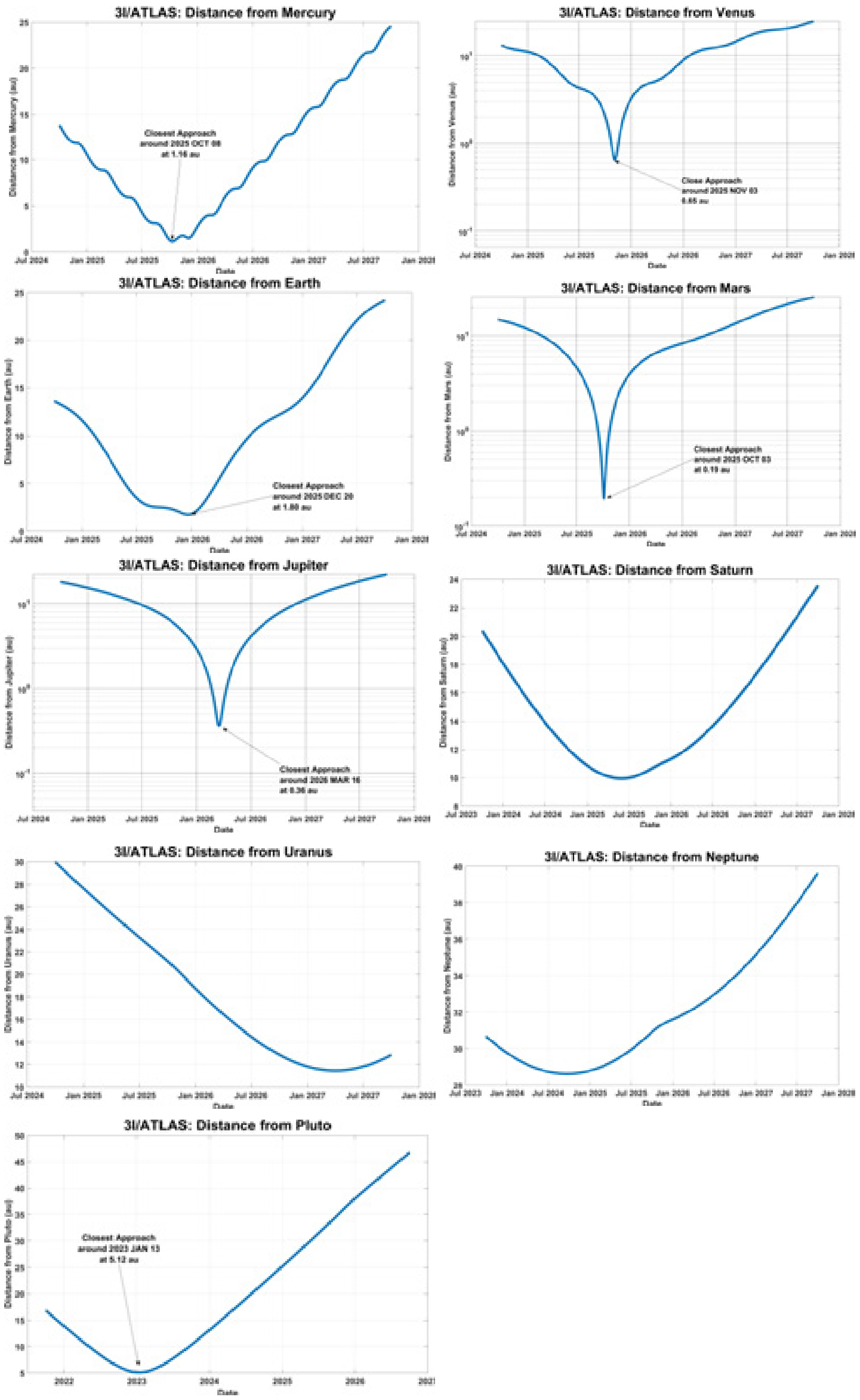

**Figure 3** Evolution of distance of 3I/ATLAS to all the planets and Pluto.

A speed of ~ 5 km s$^{-1}$ is about equivalent to the speed of an intercontinental ballistic missile, which are generally rocket-propelled. Thus assuming a similar means of propulsion (chemical), the object 3I/ATLAS could quite easily release probes of the same size, that would reach planets of interest.

Alternatively, 3I/ATLAS might intend to slow down and settle either into a heliocentric bound orbit, or a Jupiter bound one. There is good reason why it might choose a relatively low perihelion (i.e. 1.35 au on 29 October 2025), since it would then be able to exploit the 'Oberth effect' and apply all its thrust at this perihelion (Solar Oberth), or on the other hand why it would select a trajectory which swings close by Jupiter (Jupiter Oberth). Clearly, our Sun and Jupiter are the two most massive bodies in the Solar System and therefore permit a spacecraft to capitalize most on the Oberth effect, enabling a minimum ΔV requirement from the spacecraft's propulsion system.[31]

As a reminder, an Oberth manoeuvre is one where thrust of a spacecraft is applied at its maximum orbital speed, namely at periapsis,[31] so as to maximise the resulting change in kinetic energy. This applies both to accelerating to achieve Solar System escape, or alternatively to slow down from a high speed (a 'reverse Oberth manoeuvre').

Examining Figure 4, we observe that the optimal arrival dates for such an intercept visit either by the object itself, or alternatively a probe or weapon sent by it, will be from 21$^{st}$ November 2025 to 5$^{th}$ December 2025, and so this is a testable prediction of the veracity of this hypothesis.

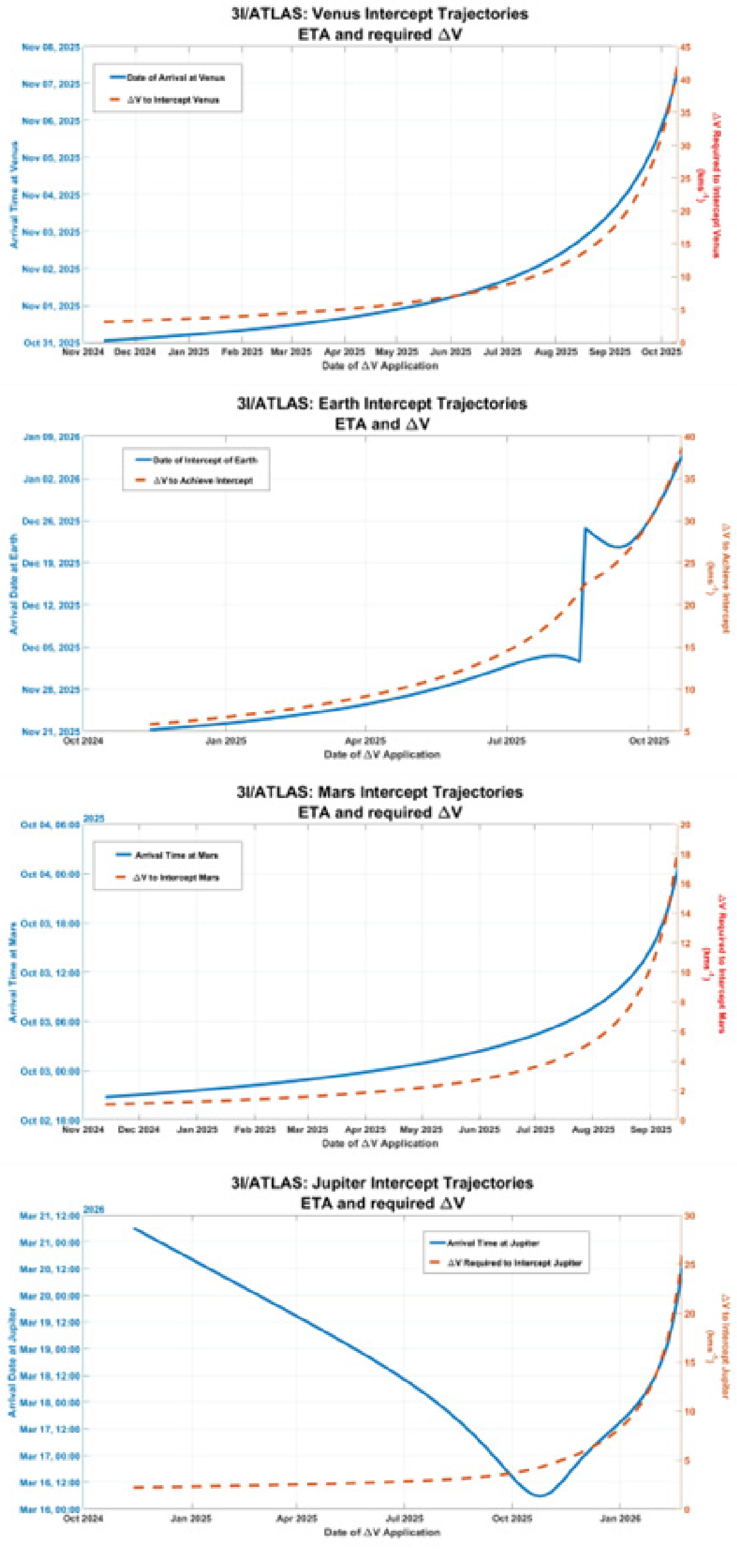

**Figure 4** Optimal ΔVs for the 4 planets in question (right vertical axis) and also the expected time of arrival (left), vs the date of ΔV application.

The intercept option would possibly indicate a malign intent, let us now consider in more detail the possibility that 3I/ATLAS wishes to rendezvous with Earth (see Figure 5). We find that this option is indeed available to 3I/ATLAS and the total ΔV (intercept + rendezvous) is lower the earlier the date of its application. Should 3I/ATLAS wish to apply this ΔV clandestinely at a low perihelion and at a low solar elongation, the sooner the execution of this initial ΔV in the window of opportunity, the better. Figure 6 shows that there is a minimum rendezvous ΔV at Earth for an arrival date around March of 2026.

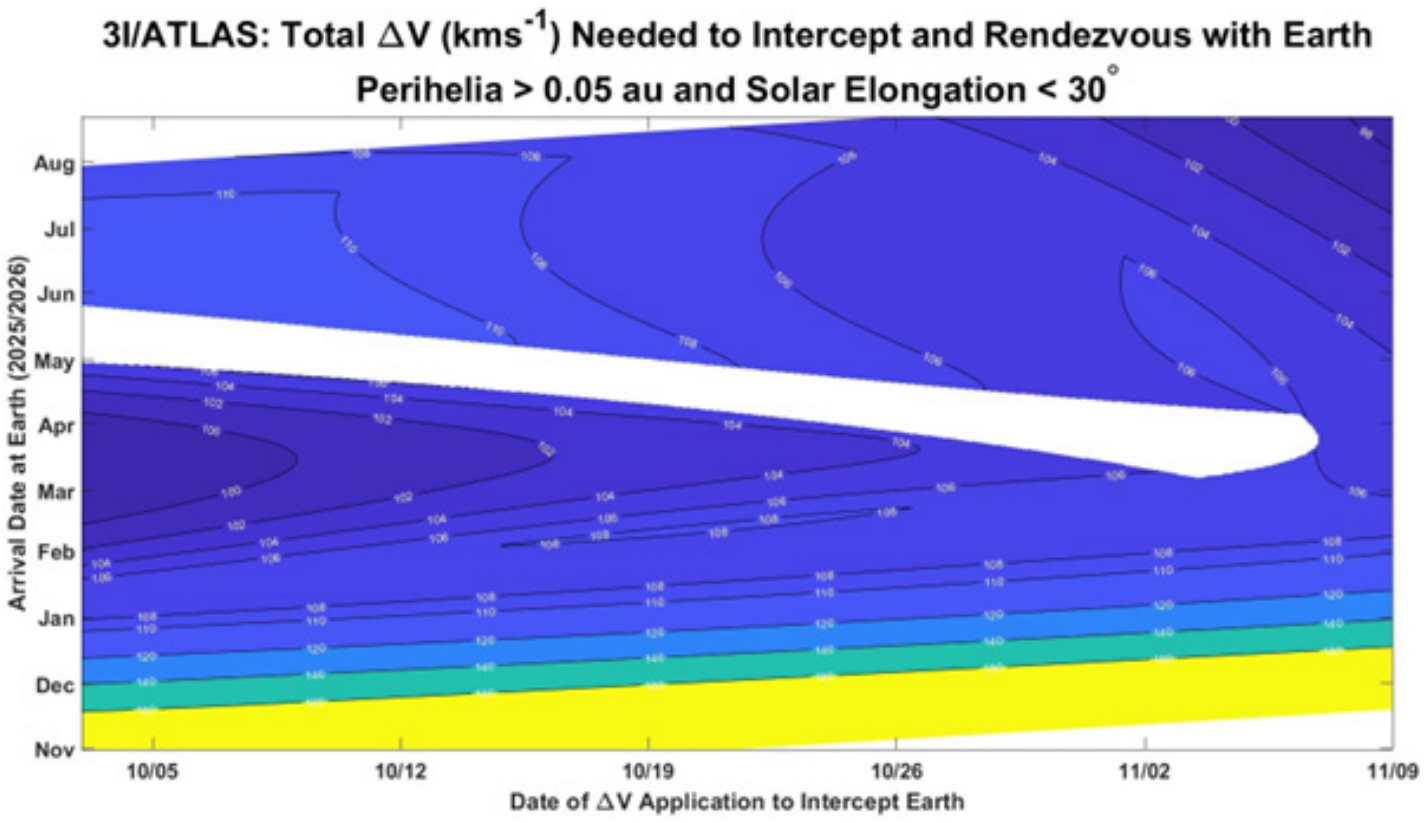

**Figure 5** Thrust ΔV colour contours to enable 3I/ATLAS to intercept the Earth at low solar elongation, given its delivery date (in 2025) and arrival date. Blank areas indicate low perihelia.

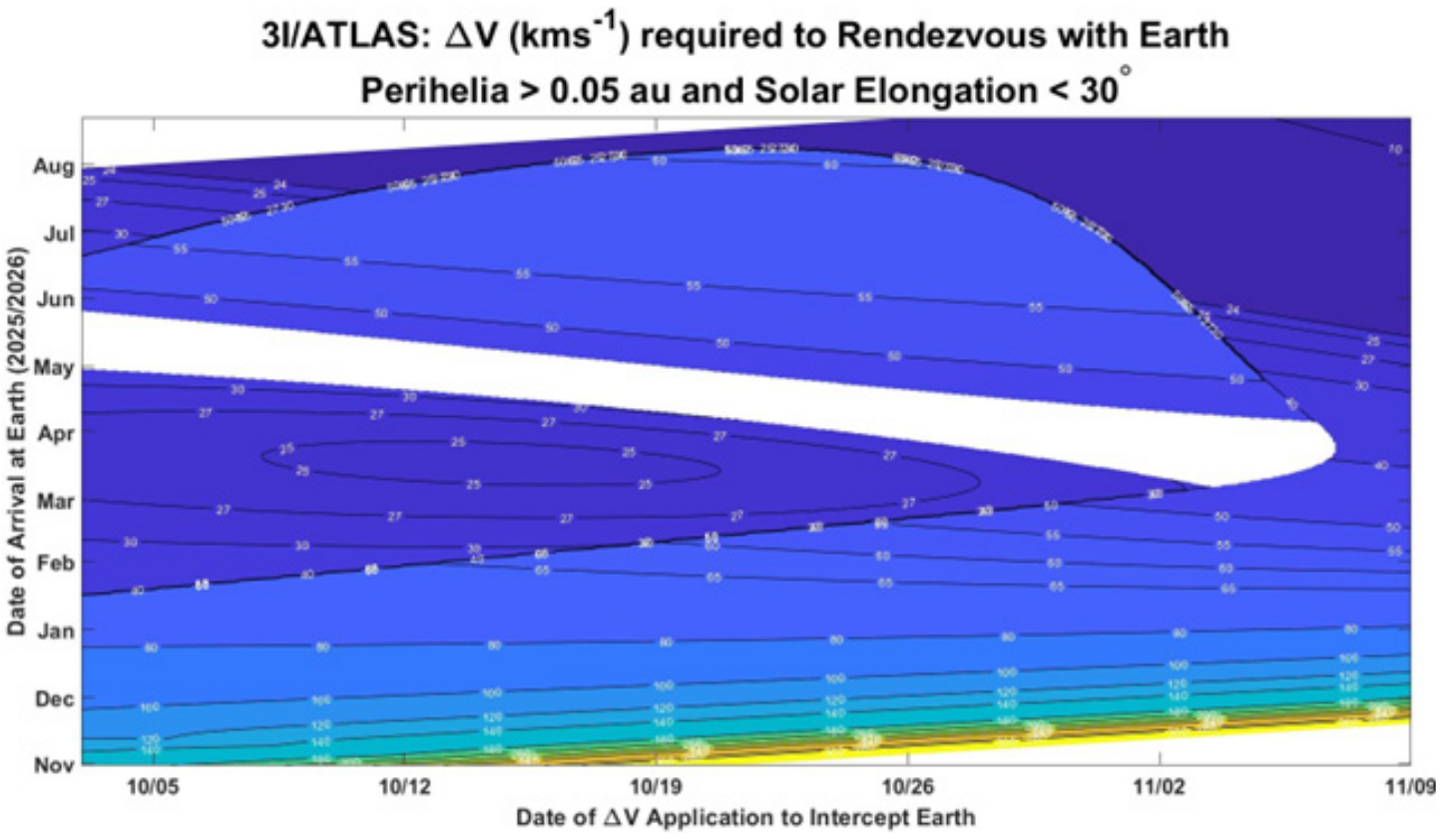

**Figure 6** Thrust ΔV colour contours to enable 3I/ATLAS to rendezvous with Earth after a low solar elongation, given the intercept delivery date (in 2025) and arrival date. Blank areas indicate low perihelia.

As mentioned, there is the chance that 3I/ATLAS will conduct a Jupiter Oberth, as described in Figure 7. Since the purpose of this Oberth would be to match velocities (rendezvous) with Jupiter, this involves a delivery of thrust both firstly to intercept Jupiter and then again to slow down into a Jupiter parking orbit. Figure 8 reveals that a ΔV of at least ~ 20 km s$^{-1}$ would be necessary upon arrival at Jupiter, assuming a perijove altitude at 0.05 Jupiter radii.

## Non-gravitational accelerations

So far we have addressed impulsive (high thrust) ΔV manoeuvres available to 3I/ATLAS, but what about low thrust manoeuvres? It is possible using a NOMAD,[21] REBOUND[32,33] and SPICE[34,35] software application, developed specifically for the purpose, to determine the minimum overall magnitude of non-gravitational acceleration components ($A_1$, $A_2$, $A_3$) radial, transverse and perpendicular to the orbital plane respectively,[30] for 3I/ATLAS to intercept Mars or Jupiter (see Table 3). Note that these accelerations are calculated assuming a start of simulation on 8$^{th}$ July 2025, and are normalized at 1 au.

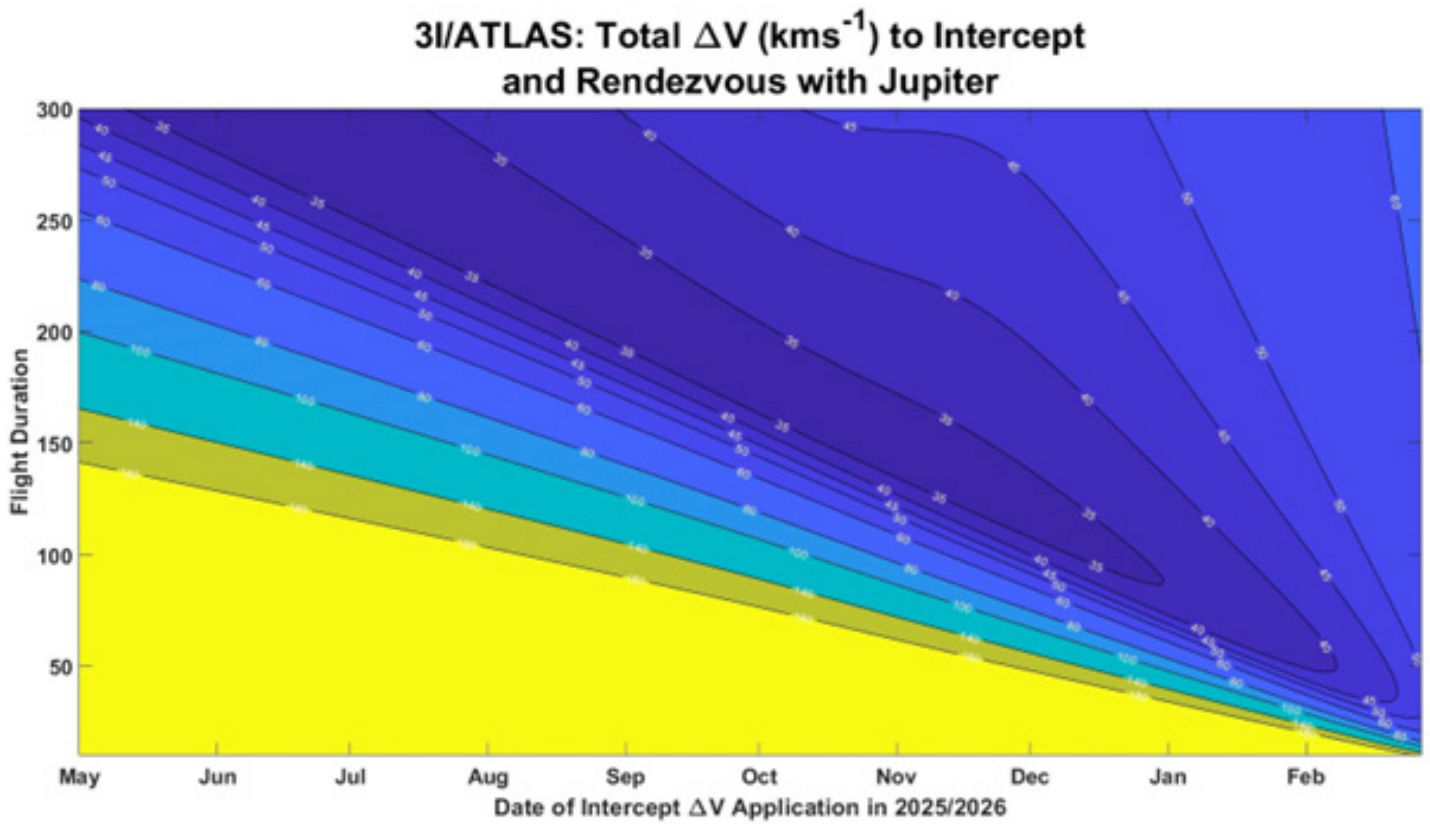

**Figure 7** ΔV colour contours to enable 3I/ATLAS to intercept and rendezvous (i.e. stay in a bound orbit) with Jupiter. Dates on the x-axis are in 2025/2026.

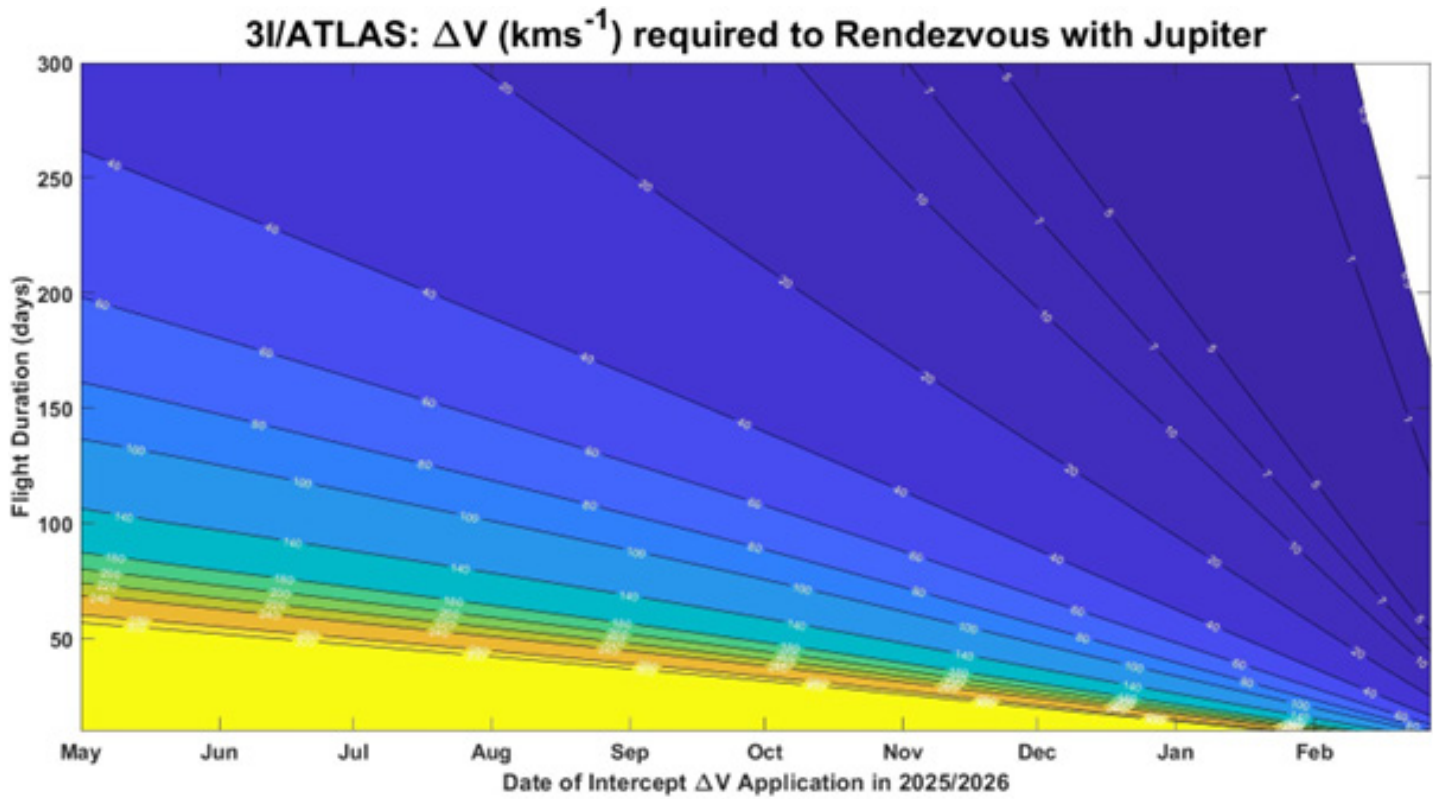

**Figure 8** ΔV colour contours to enable 3I/ATLAS to rendezvous (i.e. stay in a bound orbit) with Jupiter. Dates on the x-axis are in 2025/2026.

We find that the $A_1$ radial component for a Jupiter intercept is positive, suggesting this could be achieved by a photonic (solar) sail. If we take the magnitude for Jupiter, we have $A \sim 5.85 \times 10^{-5}$au day$^{-2}$ normalized at 1 au (equivalent to 1.17 mm/s$^2$). Assuming a perfectly reflective sail, a sail areal density of $\sigma$, a critical acceleration of $a_c$, and further that the angle the sail-normal makes with the anti-radial direction is 0°, which is the upper extreme, we have at 1 au from the Sun[37,38]:

$$\sigma/(g/m^2) <= \frac{(9.08/\mu N)}{(a_c/(mm/s^2))} \qquad (2)$$

Inserting $a_c$ = 1.17mm/s$^2$ leads to an upper limit on $\sigma$ < 7.8 g/m$^2$. This is typical of the areal density of sails humanity has developed. For example a sheet of material of mass density 7,800,000 g/m$^3$ (not far off iron for example) and with thickness 1 μm would have the required areal density. A circular sail of radius ~ 10 km (the current estimate of the size of 3I/ATLAS based on no cometary activity) would have a mass of ~ 4.8 × 10$^6$ kg. The lightness number of the Solar Sail $\lambda$ (independent of Sun-distance) is defined as the ratio of force from solar radiation pressure to that of gravity.

We find that $\lambda \gtrsim 0.20$. As of writing, the Minor Planet Center does not provide estimates of the non-gravitational accelerations for this object.

## Jupiter oberth and aero-capture

While 3I/ATLAS might be a circular sail, another option, with the advantage of a high degree of symmetry is a sphere. To maintain rigidity and strength as well as a very low areal mass density led previous researchers to propose interstellar spherical sails made of either graphene,[39-41] or Aerographite.[42,43] For this discussion we shall assume multi-layer graphene, with an areal mass density of $0.758\times10^{-7}$ kg m$^{-2}$, and an absorptivity of 0.02. To achieve 3I/ATLAS's observed albedo of 0.04 (i.e. an absorptivity of 0.96) means at least 96 layers. With a radius of 10 km, 3I/ATLAS's outer layer masses at least 91.4 tonnes. Assuming the normalised acceleration advised above, 3I/ATLAS acting as an absorbing Sail, masses at most 1,160 tonnes. Thus it can contain multiple internal layers to brace it against outside forces as well as significant payload, like sub-probes and sensor equipment. The implied cross-sectional mass-density is 0.00369 kg m$^{-2}$.

At Jupiter, 3I/ATLAS has several options one of which is Aero-Capture into a Highly Elliptical Jupiter Orbit (HEJO). Using OITS to model the orbit gives a Hyperbolic Excess of 65.5913 km s$^{-1}$ relative to Jupiter's centre.

Assuming the current physical parameters for Jupiter[44] (See Table 4) 3I/ATLAS will be moving at 88.33 km s$^{-1}$ when encountering the atmosphere at an altitude of ~ 899 km above the 1 bar reference level. Jupiter sidereal rotation period of 35,730 seconds gives an Equatorial rotational speed of 12.729 km s$^{-1}$ at that altitude. Therefore the Entry Speed is 75.6 km s$^{-1}$ relative to the atmosphere.

**Table 4.** Pertinent Jupiter parameters from NTRS - NASA Technical Report Service (2021)

| Jupiter | Label | Units | Value |
|---|---|---|---|
| **Gravitational Parameter** | GM | km$^3$ s$^{-2}$ | $1.27\times10^8$ |
| **Mean Equatorial Radius** | Re | km | 71492 |
| **Mean Polar Radius** | Rp | km | 66854 |
| **J2 Harmonic** | J2 | N/A | 0.014736 |
| **Period** | | s | 35730 |

For a capture orbit with an Apo-Jove of 527 Jupiter radii, the Peri-Jove speed needs to be 58.74 km s$^{-1}$, or 46 km s$^{-1}$ relative to the atmosphere. These are just first pass estimates, to give some idea of the re-entry conditions. Assuming a constant braking in a path equal to one Jupiter radius in length ($71.5\times10^6$ m), the deceleration for Aero-Capture is just 25 m s$^{-2}$. As 3I/ATLAS is above circular orbital velocity for the whole event, the centrifugal acceleration it experiences is counter to gravity, unlike a landing re-entry which must maintain positive lift to maximize braking distance at high altitude.

With such a low areal mass-density of $3.69\times10^{-3}$ kg m$^{-2}$, the energy dissipation is 7.01 kW m$^{-2}$ – high, but not the multi-megawatt levels experienced during an Apollo mission Lunar re-entry to Earth.[45] Additionally the total heat load is 6.64 MJ m$^{-2}$, much lower than the 426.5 MJ m$^{-2}$ experienced by Apollo. Once decelerated into orbit, conceivably 3I/ATLAS can interact with Jupiter's Galilean moons and magnetic field to maneuver without use of propellant.

Covert reconnaissance of an inhabited star system is described in some detail in Stanislaw Lem's influential fictional critique of SETI, "Fiasco".[46] In the novel Earth vehicle "Hermes" approaches an inhabited star system and adopts the guise of a parabolic comet, applying an artificial crust and ejecting gases. Later it uses close flybys of gas giants to decelerate and obscure its propulsive manoeuvering. "Hermes" adopts this cautiously covert entry into a star system due to observing possible hostile activities remotely spread across the target system, centred on the inhabited planet.

How its orbit changes during and after this loitering period needs to be studied carefully in the coming months.

## Discussion

We have proposed a testable hypothesis, that 3I/ATLAS is technological, and have demonstrated various lines of evidence to substantiate this hypothesis (see Table 1). The orbital path of 3I/ATLAS has some very unlikely combination of characteristics, which could quite easily have been simple coincidence, as extremely strange as that ostensibly appears. The propensity for the human brain to see patterns in what is actually random scatter is well known.

At the heart of this, is a question any self-respecting scientist will have had to address at some point in their career: "is an outlier of a sample a consequence of expected random fluctuation, or is there ultimately a sound reason for its observed discrepancy?" A sensible answer to this hinges largely on the size of the sample in question, and it should be noted that for interstellar objects we have a sample size of only 3, therefore rendering an attempt to draw inferences from what is observed rather problematic.

However we will have centre stage as 3I/ATLAS ventures through our Solar System, except for around its perihelion, and our telescopes currently trained on this object should show any anomalies indicative of technology in the coming months, though these may only become apparent when 3I/ATLAS has passed perihelion. As already discussed, a visitor to Earth around the end of November to the beginning of December 2025, whatever form that might take, would clearly support our supposition, and furthermore the measurement of significant non-gravitational accelerations (Table 3) would be a huge find.

## Conclusion

We strongly emphasize that this paper is largely a pedagogical exercise, with interesting discoveries and strange serendipities, worthy of a record in the scientific literature. By far the most likely outcome will be that 3I/ATLAS is a completely natural interstellar object, probably a comet, and the authors await the astronomical data to support this likely origin.

Nevertheless when viewed from an open-minded and unprejudiced perspective, these investigations have revealed many compelling insights into the possibility that 3I/ATLAS is technological, and moreover the calculations presented here are useful even if the interstellar object ends up being a comet like 2I/

Borisov because they could be applied to future detections of interstellar objects by the Vera C. Rubin observatory over the coming decade.

## Acknowledgments

Avi Loeb was supported in part by Harvard's Black Hole Initiative and the Galileo Project. Adam Hibberd and Adam Crowl worked voluntarily for the Initiative for Interstellar Studies.

## Conflicts of interest

Authors declare that there is no conflict of interest.

## References

1. Alvarez-Candal A, Rizos JL, Lara LM, et al. X-SHOOTER Spectrum of Comet C/2025 N1: Insights into a Distant Interstellar Visitor. arXiv; 2025.
2. Bolin BT, Belyakov M, Fremling C, et al. Interstellar comet 3I/ATLAS: discovery and physical description. arXiv; 2025.
3. Hopkins MJ, Bannister MT, Lintott C. Predicting Interstellar Object Chemodynamics with Gaia. arXiv; 2024.
4. Hopkins MJ, Dorsey RC, Forbes JC, et al. From a Different Star: 3I/ATLAS in the context of the Otautahi-Oxford interstellar object population model. arXiv; 2025.
5. Kakharov S, Loeb A. Galactic Trajectories of Interstellar Objects 1I/'Oumuamua, 2I/Borisov, and 3I/Atlas. arXiv; 2025.
6. Loeb A. Comment on "Discovery and Preliminary Characterization of a Third Interstellar Object: 3I/ATLAS." arXiv; 2025.
7. Loeb A. 3I/ATLAS is Smaller or Rarer than It Looks. Research Notes of the AAS. 2025;9(7):178.
8. Opitom C, Snodgrass C, Jehin E, et al. Snapshot of a new interstellar comet: 3I/ATLAS has a red and featureless spectrum. arXiv; 2025.
9. Seligman DZ, Micheli M, Farnocchia D, et al. Discovery and Preliminary Characterization of a Third Interstellar Object: 3I/ATLAS. arXiv; 2025.
10. Taylor AG, Seligman DZ. The Kinematic Age of 3I/ATLAS and its Implications for Early Planet Formation. arXiv; 2025.
11. SETI Institure Home Page; 2025.
12. Bannister MT, Bhandare A, Dybczynski PA, et al. The natural history of `Oumuamua. *Nature Astronomy*. 2019;3:594-602.
13. Loeb A. On the Possibility of an Artificial Origin for 'Oumuamua. *Astrobiology*. 2022;22(12):1392.
14. Eldadi O, Tenenbaum G, Loeb A. Scientific Paradigm Resistance Evidence from the Oumuamua Debate and Cross-Disciplinary Case; 2025.
15. Micheli M, Farnocchia D, Meech K, et al. Non-gravitational acceleration in the trajectory of 1I/2017 U1 ('Oumuamua). *Nature*. 2018;559:223-226.
16. Trilling DE, Mommert M, Hora JL, et al. Spitzer Observations of Interstellar Object 1I/'Oumuamua. *The Astronomical Journal*. 2018;156(6):261.
17. Bialy S, Loeb A. Could Solar Radiation Pressure Explain 'Oumuamua's Peculiar Acceleration? *The Astrophysical Journal*. 2018;868:L1.
18. Hibberd A. Github repository for OITS; 2017.
19. Hibberd A. Intermediate Points for Missions to Interstellar Objects Using Optimum Interplanetary Trajectory Software. arXiv; 2022.
20. Hibberd A, Perakis N, Hein AM. Sending a spacecraft to interstellar comet 2I/Borisov. *Acta Astronautica*. 2021;189:584-592.
21. Le Digabel S. Algorithm 909: NOMAD: Nonlinear Optimization with the MADS Algorithm. *ACM Transactions on Mathematical Software (TOMS)*. 2011;37(4):1-15.
22. Schlueter M, Egea J, Banga J. Extended ant colony optimization for non-convex mixed integer nonlinear programming. *Computers and Operations Research*. 2009;36(7):2217-2229.
23. Schlueter M, Erb S, Gerdts M, et al. MIDACO on MINLP space applications. *Advances in Space Research*. 2013;51(7):1116-1131.
24. Schlueter M, Gerdts M. The oracle penalty method. *Journal of Global Optimization*. 2010;47:293-325.
25. Hein AM, Perakis N, Eubanks TM, et al. Project Lyra: Sending a spacecraft to 1I/'Oumuamua (former A/2017 U1), the interstellar asteroid. *Acta Astronaut*. 2019;161:552-561.
26. Hein AM, Eubanks TM, Lingam M, et al. Interstellar Now! Missions to Explore Nearby Interstellar Objects. *Adv Space Res*. 2022;69(1):402-414.
27. Hibberd A. Project Lyra: Another possible trajectory to 1I/'Oumuamua. *Acta Astronautica*. 2023;211:431-434.
28. Hibberd A. Project Lyra: The Way to Go and the Launcher to Get There. arXiv; 2023.
29. Hibberd A, Hein AM. Project Lyra: Catching 1I/'Oumuamua–Using Nuclear Thermal Rockets. *Acta Astronautica*. 2021;179:594-603.
30. Hibberd A, Hein AM, Eubanks TM. Project Lyra: Catching 1I/'Oumuamua – Mission opportunities after 2024. *Acta Astronautica*. 2020;170:136-144.
31. Blanco PR, Mungan CE. High-speed escape from a circular orbit. *American Journal of Physics*. 2021;89:72-79.
32. Rein H, Liu SF. REBOUND: an open-source multi-purpose N-body code for collisional dynamics. *Astronomy & Astrophysics*. 2012;537:A128.
33. Rein H, Spiegel DS. ias15: a fast, adaptive, high-order integrator for gravitational dynamics, accurate to machine precision over a billion orbits. *Monthly Notices of the Royal Astronomical Society*. 2015;446(2):1424-1437.
34. Acton C, Bachman N, Semenov B, et al. A look towards the future in the handling of space science mission geometry. *Planetary and Space Science*. 2018;150:9-12.
35. Acton CH. Ancillary data services of NASA's Navigation and Ancillary Information Facility. *Planetary and Space Science*. 1996;44(1):65-70.
36. Marsden BG, Sekanina Z, Yeomans DK. Comets and nongravitational forces. V. Astronomical Journal. 1973;78:211.
37. Les Johnson. Solar Sail Propulsion – Ready for SmallSat Mission Implementation; 2024. 20 p.
38. Maurya MK, Sultana R, Baghel RS, et al. Payload and sail loading dependence study of design sensitivity function and characteristic acceleration of solar sail. *Indian Journal of Physics*. 2023;97:2591-2598.
39. Heller R, Hippke M. Deceleration of High-velocity Interstellar Photon Sails into Bound Orbits at α Centauri. *Astrophysical Journal Letters*. 2017;835(2):L32.
40. Heller R, Hippke M, Kervella P. Optimized Trajectories to the Nearest Stars Using Lightweight High-velocity Photon Sails. *Astronomical Journal*. 2017;154(3):115.

41. Matloff GL. Journal of the British Interplanetary Society. 2013;66:377.

42. Heller R, Anglada-Escude G, Hippke M, et al. Low-cost precursor of an interstellar misión. *Astronomy & Astrophysics*. 2020;641:A45.

43. Karlapp J, Heller R, Tajmar M. Ultrafast transfer of low-mass payloads to Mars and beyond using aerographite solar sails. *Acta Astronautica*. 2024;219:889-885.

44. Justh HL, Hoffman J, Dwyer AM. Jupiter Global Reference Atmospheric Model (Jupiter-GRAM): User Guide; 2021.

45. Pavlosky AE, St. leger LG. Apollo experience report: Thermal protection subsystem; 2013.

46. Lem S. Fiasco (Harcourt Brace Jovanovich); 1987.